\documentclass[%
 reprint,
 longbibliography,
 amsmath,
 amssymb,
 aps,
 pre,
]{revtex4-1}

\usepackage{dcolumn}
\usepackage{bm}
\usepackage{hyperref}

\usepackage{amssymb}
\usepackage{amsmath}
\usepackage{cases}
\usepackage{graphicx,color}
\definecolor{r}{rgb}{1,0,0}   
\definecolor{g}{rgb}{0,1,0}   
\definecolor{b}{rgb}{0,0,1}
\definecolor{purple}{rgb}{0.808,0.454,0.718}

\begin{document}


\title{The Hindered Settling Function at Low Re Has Two Branches}


\author{T. A. Brzinski III and D. J. Durian}
\affiliation{Department of Physics \& Astronomy, University of Pennsylvania, Philadelphia PA 19104, USA}
\date{\today}

\begin{abstract}
We analyze hindered settling speed versus volume fraction $\phi$ for dispersions of monodisperse spherical particles sedimenting under gravity, using data from 15 different studies drawn from the literature, as well as 12 measurements of our own.
We discuss and analyze the results in terms of popular empirical forms for the hindered settling function, and compare to the known limiting behaviors.
A significant finding is that the data fall onto two distinct branches, both of which are well-described by a hindered settling function of the Richardson-Zaki form $H(\phi)=(1-\phi)^n$ but with different exponents: $n=5.6\pm0.1$ for Brownian systems with P\'eclet number ${\rm Pe}<{\rm Pe}_c$, and $n=4.48\pm0.04$ for non-Brownian systems with ${\rm Pe}>{\rm Pe}_c$.
The crossover P\'eclet number is ${\rm Pe}_c\approx10^8$, which is surprisingly large. 
\end{abstract}



\maketitle



\section{Introduction}

When solid particles are dispersed into a fluid, there is inevitably a mass-density mismatch.
Therefore sedimentation under the influence of gravity happens generically in all suspensions, and understanding and controlling this behavior is a widespread issue of both pure and applied interest \cite{Barnea73, Garside77, DavisAcrivos, PhilipseCOCIS97, Burger01, GuazzelliMorris, PiazzaRPP14}. 
For example: in geophysical sciences the physics of sedimentation controls sediment deposition and transport~\cite{gyr2003sedimentation,TurbidGeo1,TurbidGeo2}; in industry, sedimentation and decanting has long been utilized as a means of separating solids from liquid solvents -- notably, this is a key process in most wine-making techniques, and thus dates back thousands of years, but also plays an important role in modern industries such as petroleum processing and nanotechnology.

\begin{figure}[ht] 
\includegraphics[width=\columnwidth]{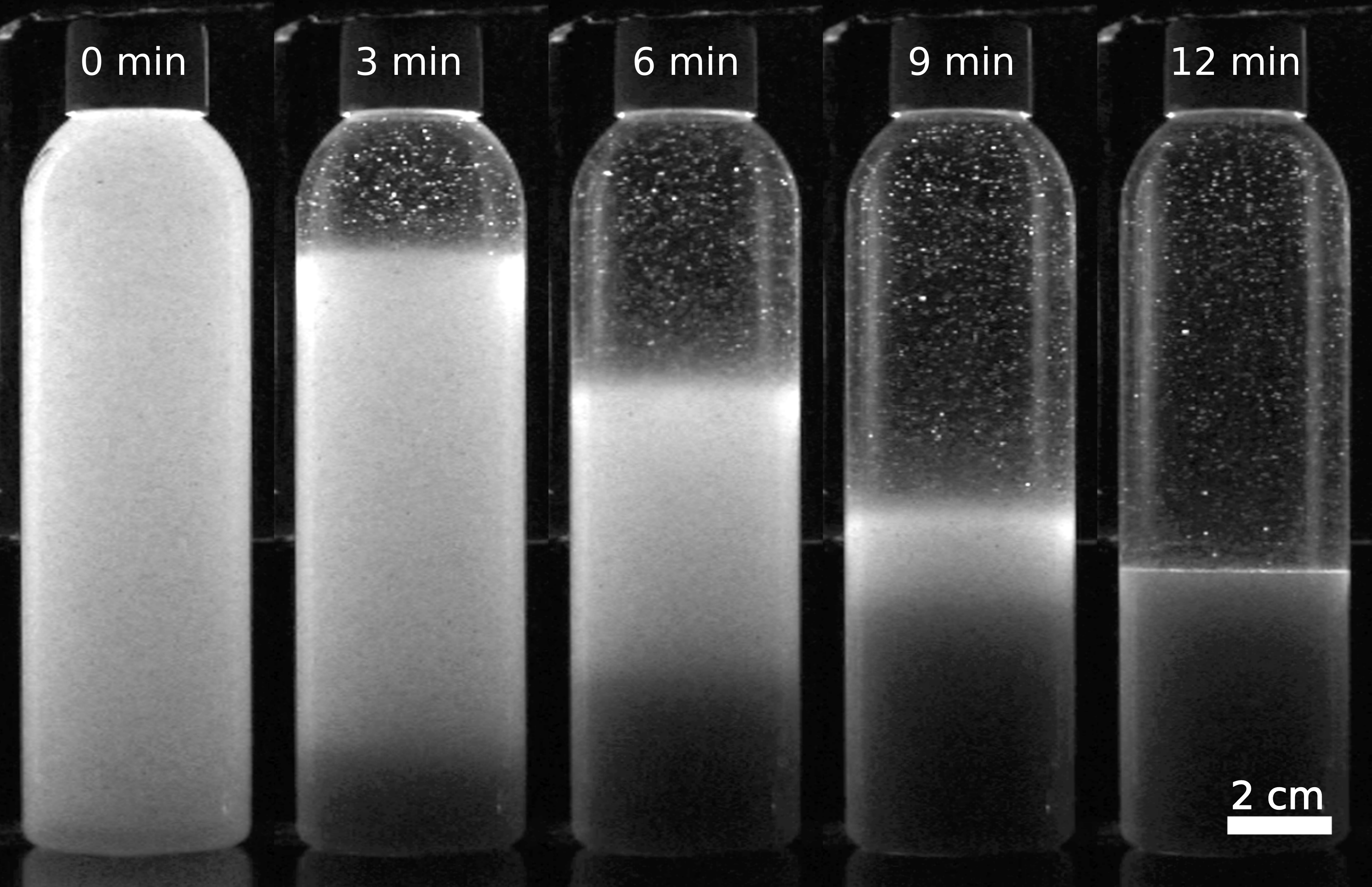}
\caption{A time series of photographs for non-Brownian $d=365~\mu$m glass spheres with initial volume fraction $\phi=0.21$ sedimenting in an aqueous glycerol solution. The sample is illuminated from both sides, so the only light which reaches the camera is that which is scattered at close to 90-degrees.  Thus the dark regions at the bottom are densely-packed sediment, the dark regions at the top are depleted of particles, and the bright regions in the middle are uniformly-dispersed grains that multiply scatter light toward the camera.}
\label{Montage}
\end{figure}

To isolate key features, researchers often focus on samples where the particle volume fraction $\phi$ is initially uniform and the container has vertical sidewalls and a fixed horizontal bottom as depicted in Fig.~\ref{Montage}.
If the particles are monodisperse, then the sedimentation rate is constant and hence $\phi$ remains uniform throughout the suspension.
Consequently there arise two fronts that are readily visible in Fig.~\ref{Montage}: a supernatant-suspension front that move downwards from the top at the sedimentation speed $v$ and a sediment-suspension front that moves upwards from the bottom as particles deposit out.
For large non-Brownian particles, sedimentation stops when the two fronts meet and all particles are packed at rest with volume fraction $\phi_c$ at the bottom of the container.
For small Brownian particles, the initial sedimentation rate is constant but the two fronts are affected and the final state is an exponential concentration profile.
There is a wealth of interesting additional behavior concerning velocity fluctuations and the effects of particle size / shape / polydispersity / interactions as well as initial conditions, boundaries, container shape, and applied shear.

The most basic issue is to understand how the sedimentation speed $v$ varies with volume fraction $\phi$ for non-interacting monodisperse spheres in a Newtonian fluid at low Reynolds number.
At very low $\phi$, the sedimentation speed must approach the Stokes speed
\begin{equation}
	v_s=\frac{2\Delta \rho g a^2}{9\eta}
\label{eq:Stokes}
\end{equation}
for an individual grain, where $\Delta \rho = \rho_p-\rho_f$, $\rho_p$ is particle density, $\rho_f$ is fluid density, $g$ is gravitational acceleration, $a$ is particle radius, $d=2a$ is particle diameter, and $\eta$ is fluid viscosity.
At nonzero $\phi$, the mean sedimentation speed is slower due to hindering by the tortuous upward fluid flow between particles, which themselves experience significant velocity fluctuations.
This may be characterized empirically by a dimensionless ``hindered settling function'' $H(\phi)\le 1$ defined by the mean sedimentation speed via
\begin{equation}
	v=v_sH(\phi).
\label{eq:Hindered}
\end{equation}
Despite decades of research, there is still great uncertainty and conflicting reports for the form of $H(\phi)$ in the primary literature and in reviews.
This is reflected recently in Ref.~\cite{GuazzelliMorris}, which shows two data sets for $\phi \le 0.4 \approx (2/3)\phi_c$ \cite{Ham88, Nicolai95} and states that the empirical Richardson-Zaki \cite{Richardson54} form $H(\phi)=(1-\phi)^n$, with ``$n\approx 5$ most accurately represents the experimental data for small Reynolds number \ldots\ this correlation is likely to be inaccurate when approaching maximum packing."
Indeed, one of the data sets \cite{Ham88} has a very small range and the other \cite{Nicolai95} shows a clear systematic deviation from the plotted Richardson-Zaki function.  Furthermore, we have encountered different values of $n$ ranging from about 4--7 quoted by different authors as the accepted value.

In this paper, we significantly clarify the form of $H(\phi)$ across the full range of volume fractions, $0 \le \phi \le \phi_c$, where the Reynolds number is small and the P\'eclet number varies over fifteen orders of magnitude.
We begin by discussing expectations for the functional form of $H(\phi)$ versus $\phi$ based on prior theory and empirical fits to data.
Next we gather hindered settling data from fifteen sources in the literature, and describe our own measurement procedures and results.  
Finally we collate all data into plots, as well as a Supplemental data file \cite{supp}, and compare with common forms of $H(\phi)$ versus $\phi$, including a cumulant expansion that we propose.

\section{Expectations}

We restrict attention to non-interacting (i.e.\ ``hard") monodisperse spherical particles of diameter $d=2a$ and density $\rho_p$ at constant uniform volume fraction $\phi$ in a Newtonian fluid of viscosity $\eta$ and density $\rho_f$.
Different classes of behavior are potentially controlled by the dimensionless Reynolds and P\'eclet numbers, respectively defined and evaluated based on the Stokes speed $v_s$ of Eq.~(\ref{eq:Stokes}) as
\begin{eqnarray}
	{\rm Re} &\equiv& \rho_f v_s a/\eta = \frac{2\rho_f \Delta\rho g a^3}{9\eta^2}, \label{Re} \\
	{\rm Pe} &\equiv& v_s a/D_o = \frac{4\pi\Delta\rho g a^4}{3kT}. \label{Pe}
\end{eqnarray}
These are single-particle quantities that describe the system constituents independent of particle volume fraction, $\phi$, per usual practice.
The Reynolds number indicates the importance of inertial to viscous forces.
In this work we only consider systems with small ${\rm Re}$, where inertial effects can be neglected such that $v\rightarrow v_s$ in the $\phi\rightarrow 0$ dilute limit.
The P\'eclet number indicates the importance of flow relative to thermal diffusion.
As standard, it is defined by the single-sphere diffusivity $D_o=k_BT/6\pi\eta a$ \cite{GuazzelliMorris, Ackerson07, Chaikin92}.
Other choices are possible, e.g. the zero- or long-wavelength diffusivity of the suspension; however, these are $\phi$-dependent and are based on the actual collective behavior rather than just on the individual constituents of the system.
The non-Brownian limit, where diffusion can be neglected, is ${\rm Pe}\rightarrow\infty$.
Here, we consider settling data for both small Pe (Brownian) and large Pe (non-Brownian).
Note that both Re and Pe increase very rapidly with particle radius.

Naturally there are many highly-cited reviews of sedimentation.
For example, the recent book by Guazzelli \& Morris \cite{GuazzelliMorris} beautifully introduces general fluid dynamics topics regarding suspensions, and has a chapter on sedimentation that equivocally recommends $H(\phi) \approx (1-\phi)^n$ with $n\approx 5$ as noted above.
They also refer the reader to specialized reviews by Davis \& Acrivos \cite{DavisAcrivos} and by Guazzelli \& Hinch \cite{GuazzelliHinch11}.
These two reviews emphasize topics other than hindered settling, but both briefly mention $n\approx 5$ and note how it differs from Batchelor's famous calculation of $H(\phi)=1-6.55\phi+\mathcal O(\phi^2)$ \cite{Batchelor72}.
Davis \& Acrivos cite reviews of hindered settling by Garside \& Al-Dibouni \cite{Garside77} and by Barnea \& Mizrahi \cite{Barnea73}.
These in turn display literally dozens of empirical hindered settling functions and fitting parameters, with a primary view toward non-negligible Reynolds number.
For small Re, Davis \& Acrivos state that data generally fall between Richardson-Zaki with $n=5.1$ and the empirical form $H(\phi)=(1-\phi)/\{(1+\phi^{1/3})\exp[(5\phi/3)/(1-\phi)]\}$ recommended by Barnea \& Mizrahi [misquoted by Davis \& Acrivos as $H(\phi)=(1-\phi)^2/\ldots$].
However, a data plot illustrating this statement is not shown in either review.

In terms of theory, there are a few well-known works of particular note.
In 1972 Batchelor predicted $H(\phi)=1-6.55\phi+\mathcal O(\phi^2)$ \cite{Batchelor72}.
In 1988 Brady \& Durlofsky predicted $H(\phi) = (1-\phi)^3/(1+2\phi)=1-5\phi+\mathcal O(\phi^2)$ \cite{Brady88}.
Shortly thereafter Ladd reported simulation results for sedimentation with~\cite{LaddJCP90} and without~\cite{LaddPF93} Brownian motion.
And in 2000 Snabre \& Mills predicted $H(\phi) = (1-\phi)/[1+(b-1)\phi/(1-\phi)^3] = 1-b\phi+\mathcal O(\phi^2)$ \cite{Snabre00}.
Here, $b$ is an unknown parameter that ``reflects the angular dispersion of the fluid streamlines against the vertical direction" \cite{Snabre00}; $b=5.6$ is the authors' recommended value.
In 2011 Ref.~\cite{Gilleland11}, which concerns charged particles, stated ``The only approximate theory recognizing both hydrodynamic interactions and the equilibrium microstructure of disordered dispersions was formulated by Brady \& Durlofsky (1988), with pairwise additive hydrodynamics in the far-field approximation of Rotne-Prager."

Lastly, we may compare the speed $v=v_sH(\phi)$ of sedimentation driven by the pressure gradient $\Delta \rho g$ with the Darcy's law speed $v=Kd^2\nabla p/\eta$ of flow through a static porous medium of permeability $K$ driven by an imposed pressure gradient $\nabla p$.
Equating these speeds, and using both $\nabla p=\Delta \rho g$ plus Eq.~(\ref{eq:Stokes}), gives $H(\phi)=18K(\phi)$.
For sintered and compressed spheres at $\phi>0.53$, permeability data are well-described by the Kozeny-Carman function $K=(1-\phi)^3/(180\phi^2)$ \cite{Wong84, Torquato86, Torquato89, SalBook}; e.g.\ the value $K_c=6.3\times 10^{-4}$ for random close packing at $\phi_c=0.64$ is very well established \cite{Beavers73, Verneuil11}.
Thus we arrive at $H(\phi)=(1-\phi)^3/(10\phi^2)$ for $\phi>0.53$ and $H(0.64)=0.0114$ for additional comparison with hindered settling data.

\begin{figure}[ht]
\includegraphics[width=\columnwidth]{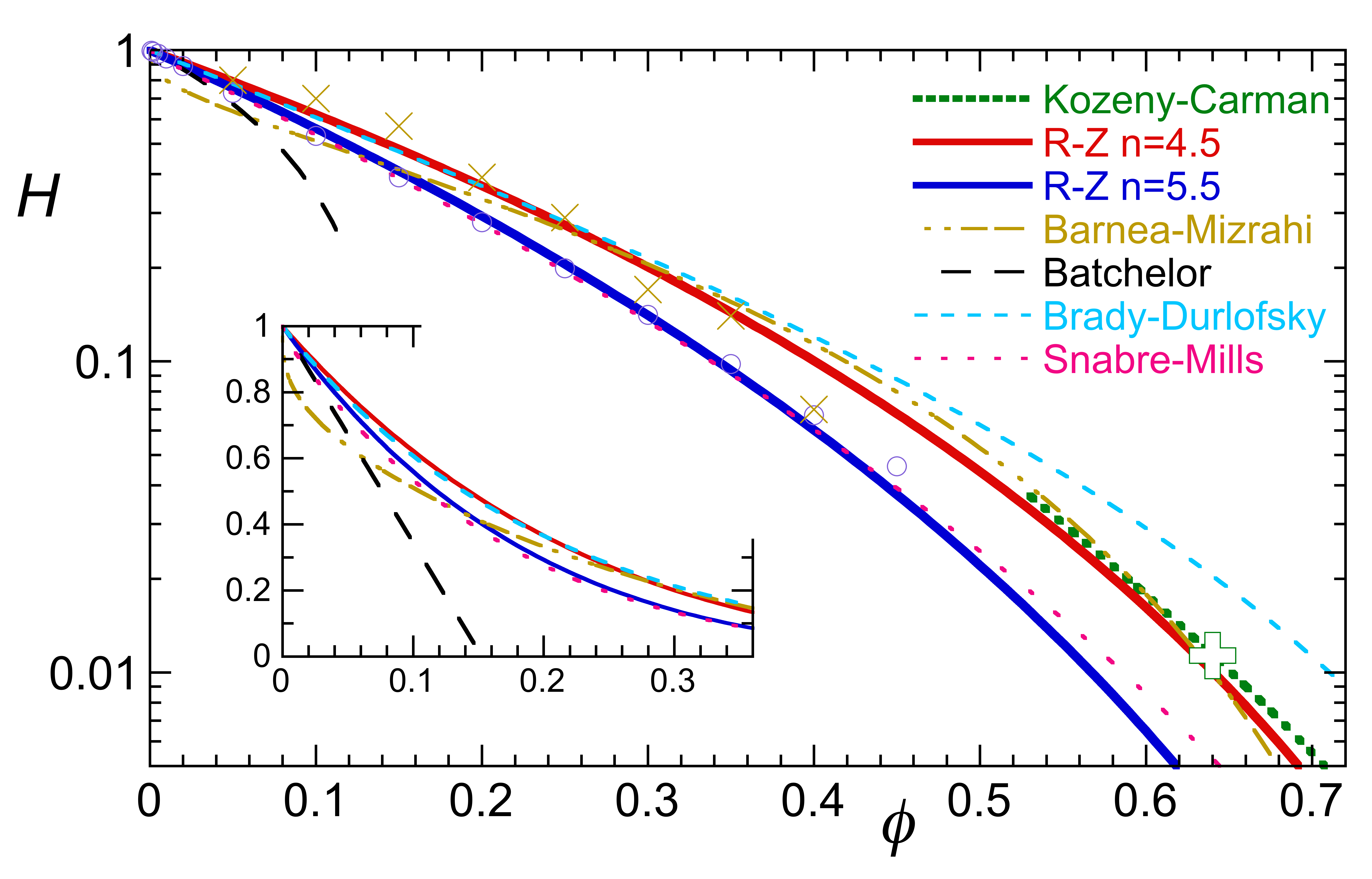}
\caption{(color online) Various expectations for the hindered settling function $H(\phi)=v/v_s$ versus volume fraction $\phi$.  See Eq.~(\ref{Hexpect}) for formulae and special values.  Crosses are experimental data \cite{Nicolai95, GuazzelliMorris};  open circles are simulation data \cite{LaddJCP90};  open plus sign is Kozeny-Carman at $\phi=0.64$.  The Snabre-Mills form is shown with the authors' recommended value of $b=5.6$ for the free parameter.}
\label{Kforms}
\end{figure}

Altogether we thus have the following expectations for the hindered settling function of non-interacting monodisperse spheres with uniform volume fraction $\phi$ in a Newtonian fluid:
\begin{widetext}
\begin{subnumcases}{\label{Hexpect} H(\phi)=}
	\label{Ko}
		1 & Stokes, $\phi=0$ \\
	\label{Kkz}
		(1-\phi)^3/(10\phi^2) & Kozeny--Carman, $\phi>0.53$ \\
	\label{Kc}
		0.0114 & Kozeny--Carman at $\phi=0.64$ \\ 
	\label{Krz}
		(1-\phi)^n & Richardson-Zaki (1954) \\
	\label{Kbm}
		(1-\phi)/\{(1+\phi^{1/3})\exp[(5\phi/3)/(1-\phi)]\} & Barnea-Mizrahi (1973) \\
	\label{Kbat}
		1-6.55\phi+\mathcal O(\phi^2) & Batchelor (1972) \\
	\label{Kbd}
		(1-\phi)^3/(1+2\phi)=1-5\phi+\mathcal O(\phi^2) & Brady-Durlofsky (1988) \\
	\label{Ksm}
		(1-\phi)/[1+(b-1)\phi/(1-\phi)^3] = 1-b\phi+\mathcal O(\phi^2) & Snabre-Mills (2000)
\end{subnumcases}	
\end{widetext}
These forms are compared in Fig.~\ref{Kforms}, along with the simulation results from Table~IV of Ref.~\cite{LaddJCP90}.
The disparity of behavior further emphasizes the need for our data compilation below.
Richardson-Zaki is shown for both $n=5.5$ and $n=4.5$, which will be seen to match well with Brownian and non-Brownian data, respectively.
Note in the figure that Brady-Durlofsky corresponds well with $n=4.5$, and hence to the non-Brownian data, for $\phi<0.35$.
Interestingly, Brady-Durlofsky exactly matches the rigorous lower bound on permeability through order $\phi$ \cite{Torquato86, Torquato89}.
Snabre-Mills is shown with the authors' recommended value of $b=5.6$; note that this form corresponds well with $n=5.5$, and hence to the {\it Brownian} data.
However, Snabre-Mills repeatedly state that their theory and the data sets they selected for comparison are for {\it non}-Brownian spheres.
Finally note how Barnea-Mizrahi dips down extremely fast, as $H(\phi)=1-\phi^{1/3}+\ldots$, and then oscillates around both Richardson-Zaki curves; hence, it does not correspond well to any of the compiled data sets.
Rather, such an initial decay is predicted for a fixed periodic array of sedimenting particles \cite{DavisAcrivos}.
By contrast $1-\beta\phi^{1/2}+\ldots$ is predicted for a fixed random array \cite{DavisAcrivos}.

\section{Prior data for hindered settling}

Based on reviews and literature search engines, we have identified a large number peer-reviewed papers with original data for settling speed versus volume fraction for relatively monodisperse hard spherical particles at small ${\rm Re}$.
For all sources except \cite{Ham90}, ${\rm Re}$ is significantly less than one.
We aimed to be exhaustive, but given the long-standing importance of sedimentation in many fields we may have inadvertently missed some available data.
The identified data sets include experiments where the settling speed was measured directly from the motion of the suspension-supernatant interface as a function of volume fraction \cite{Richardson54, Oliver61, Buscall82, Kops82, Bacri86, Davis88, Ham88, Ackerson90, Nicolai95, Ackerson07, Piazza08}, as well as experiments where the volume fraction was measured as a function of fluidization speed \cite{Richardson54, Ham90, Chaikin92}.
Note that Richardson \& Zaki did both \cite{Richardson54}, as did Ham \& Homsy sequentially \cite{Ham88, Ham90}.
Of all these papers, only Refs.~\cite{Oliver61, Kops82, Nicolai95} provide tables of data; for the rest we used commercial software to digitize the data from electronic copy.
A summary of system parameters for the various sources is given in Table~\ref{DataComp}.
Note that Pe varies over a huge range, from $10^{-4}$ (Brownian) to $10^{12}$ (non-Brownian).

\begin{table*}[ht]
\caption{\label{DataComp} System parameters for all sources, sorted by P\'eclet number.  Method is denoted as settling (s) or fluidization (f).  Type denotes Brownian (B) or non-Brownian (n-B) as assigned based on Figures~\ref{Hnonbrownian}-\ref{Hbrownian}.  The Reynolds and P\'eclet numbers were calculated from Eqs.~(\ref{Re},\ref{Pe}) using the tabulated fluid viscosities $\eta$, fluid $\rho_f$ and particle $\rho_p$ densities, and particle diameters $d=2a$.  All samples have a relatively small degree of polydispersity.}
\begin{ruledtabular}
\small{\begin{tabular}{lccccccc}
Source (method) & $\eta$ (g/cm-s) & $\rho_f$ (g/ml) & $\rho_p$ (g/ml) & $d$ ($\mu$m) & Type & Pe & Re\\ \hline
\cite{Kops82} Kops82, Table IV (s) & 0.01 & 0.78 & 1.77 & 0.13 & {\color{b} B} & 1.8E-04 & 4.4E-10\\
\cite{Piazza08} Buzzaccaro08, Fig.~8 (s) & 0.01 & 1.00 & 2.15 & 0.15 & {\color{b} B} & 4.1E-04 & 1.14E-09\\
\cite{Ackerson07} Benes07, Fig.~1 (s) & 0.01 & 1.00 & 1.05 & 2 -- 40 & {\color{b} B} & 0.50 -- 8.1E+04 & 1.1E-7 -- 8.7E-4\\
\cite{Ackerson90} Paulin90, Fig.~3a (s) & 0.023 & 0.93 & 1.19 & 0.99 & {\color{b} B} & 0.16 & 1.2E-08\\
\cite{Buscall82} Buscall82, Fig.~4 (s) & 0.01 & 1.00 & 1.05 & 3.05 & {\color{b} B} & 2.7 & 3.9E-07\\
\cite{Chaikin92} Xue92, Fig.~1 (f) & 0.01 & 1.00 & 1.05 & 31 & {\color{b} B} & 2.9E+04 & 4.1E-04\\
\cite{Bacri86} Bacri86, Fig.~2 (s) & 0.01 & 1.00 & 2.50 & 40 & {\color{b} B} & 2.4E+06 & 2.6E-02\\
\cite{Salin95} Martin95, spreadsheet (f$'$) & 0.02 & 1.00 & 2.50 & 69 & {\color{b} B} & 2.1E+07 & 3.4E-02\\
\cite{Richardson54} Richardson54, Fig.14a (s,f) & 0.015 & 1.00 & 1.06 & 217 & {\color{r} n-B} & 8.0E+07 & 6.8E-02\\
\cite{Oliver61} Oliver61, Table 3 (s) & 0.02 & 1.00 & 1.19 & 161 & {\color{r} n-B} & 8.1E+07 & 5.4E-02\\
\cite{Davis88} Davis88, Fig.~1 (s) & 0.85 & 1.02 & 2.49 & 130 & {\color{b} B} & 2.7E+08 & 1.2E-04\\
\cite{Ham90} Ham90, Fig.~3a (f) & 0.02 & 1.06 & 2.47 & 410 & {\color{r} n-B} & 2.5E+10 & 7.0\\
\cite{Ham88} Ham88, Fig.~4 (s) & 9.1 & 1.08 & 2.42 & 535 & {\color{r} n-B} & 6.9E+10 & 7.2E-05\\
\cite{Nicolai95} Nicolai95, Table 1 (s) & 13 & 1.09 & 2.53 & 788 & {\color{r} n-B} & 3.5E+11 & 1.2E-04\\
$[\text{this work}]$ Brzinski15 (s) & 2.2 & 1.24 & 2.53 & 180 -- 1000 & {\color{r} n-B} & 8.3E+08 -- 8.0E+11 & 5.1E-5 -- 8.8E-3
\end{tabular}}
\end{ruledtabular}
\end{table*}

The compiled dimensionless settling speed data are available as a Supplement to this work~\cite{supp}.
There are four columns: source, $\phi$, $H$, and error estimate $\Delta H$.
Only Ref.~\cite{Nicolai95} tabulates uncertainties.
The Ref.~\cite{Oliver61} table has three values for each $\phi$; we take the average, and use the average of all standard deviations for $\Delta H$.
For all other data sets we estimate $\Delta H$ from the root-mean-square deviation of the data from a smooth fit.
For some data sets we take $\Delta H$ to be constant; for others with large dynamic range we take it to be a constant fraction of the fitting function.  

In all cases except one we used the published data as-is.
The exception is Ref.~\cite{Oliver61}, which displays an initial decay for small volume fractions like the Barnea-Mizrahi form, with leading behavior $1-\beta\phi^{1/3}$.
This is the expectation for a fixed periodic array of particles~\cite{DavisAcrivos}.
Hence those data were excluded from the compilation, and the data for $\phi\ge0.05$ were normalized to $H(0)=1$ by the prefactor in the fitting result $v(\phi)/v_s=0.86(1-\phi)^{4.45}$.
As will be seen below, this brings it into agreement with the other non-Brownian data sets.

While circulating a draft of our compilation we were made aware of an additional Brownian data set \cite{Salin95} where the hindered settling function was deduced at high concentrations in a fluidized bed based on acoustic measurement of the concentration profile evolution after a flow-rate change.
This was part of the PhD thesis of J\'er\^ome Martin, who kindly sent a spreadsheet of his data.
The results are close to, but slightly above, the other Brownian data in our compilation and give a Richardson-Zaki exponent quoted as $n=5.35$ \cite{Salin95}.
These data are included in our plots, and in our Supplemental data file~\cite{supp}, but not in the fits discussed below.

\section{New data for hindered settling}

In addition to our meta-analysis, we conducted a series of sedimentation experiments for comparison with the compilation of published data.
We employ the most usual of the two standard methods for measuring $v(\phi)$, where the settling speed is found as the downward speed of the supernatant-suspension interface.
This approach is straightforward, and has been previously employed or described extensively~\cite{Richardson54, Oliver61, Buscall82, Kops82, Bacri86, Davis88, Ham88, Ackerson90, Nicolai95, Ackerson07, Piazza08} and reviewed~\cite{GuazzelliMorris, DavisAcrivos, GuazzelliHinch11, Garside77, Barnea73}.
Our detailed methods are as follows:

The materials properties for our samples are given in Table~\ref{Materials}.
In particular, the particles are soda-lime glass spheres (Potters Industries) with four different manufacturer-reported median diameters ranging between $d=180~\mu$m and 1~mm.
In order to remove surfactants that might trap air at the grain surfaces and effectively modify $\Delta\rho$, we used the following procedure to clean the grains: The grains were all soaked for 1-2 hours in 1~N aqueous HCl, then repeatedly rinsed with filtered deionized water until a pH strip read neutral; Between rinses, the grains were drained with a vacuum filter flask; After rinsing, the grains were dried for 24 hours in air at 350~C.
Once clean, the grains were weighed, then poured into a clear 6~oz plastic bottle (Container and Packaging Supply, part no. B335).
To these containers we added a 90~wt\% aqueous glycerol solution (see Table~\ref{Materials}), sufficient to completely fill the pore space of the granular packing.
These mixtures were then evacuated for a period of 2-7 days in order to further minimize the presence of air bubbles.
Once the sample was degassed, a 1/2~inch diameter brass sphere was added to facilitate the dispersion of the grains by manual shaking, and the bottle was overfilled with more of the degassed glycerol solution and capped so that no air remained inside the sample.

\begin{table*}[ht]
\caption{\label{Materials}  Materials properties: Median sphere diameter, range, and corresponding values for the Stokes settling speed, P\'eclet number, and Reynolds number for our experiments. The particulate material is soda-lime glass, with density $\rho_p$=2.53~g/ml.  The fluid is aqueous solution of 90\%/wt glycerol, with viscosity $\eta=2.20$~g/cm-s and density $\rho_f$=1.24~g/ml.}
\begin{ruledtabular}
\begin{tabular}{cccc}
$d$ ($\mu$m) & $v_s$ (mm/s)  & Pe & Re \\ \hline
$180\pm30$ & 0.103 & 8.34E+08	& 5.12E-05 \\
$365\pm65$ & 0.425 & 1.41E+10	& 4.27E-04 \\
$515\pm85$ & 0.847 & 5.59E+10	& 1.20E-03 \\
$1000\pm200$ & 3.192 & 7.94E+11	& 8.79E-03
\end{tabular}
\end{ruledtabular}
\end{table*}

To conduct an experiment, the sample was shaken vigorously by hand for a minute or more, until uniform to the eye.
Next, the sample was immediately placed on a lab jack inside a cardboard box which had been spray-painted matte black, and centered between long slits cut on opposing sides of the box.
A fluorescent tube light was mounted along each slit outside the box to uniformly illuminate the sample from the sides.
A small porthole on a third face of the box provided access so that the sample could be photographed with a Nikon DSLR camera. The camera was triggered at 60, 30, or 6~frames per minute, as appropriate to capture the dynamics.
Small flaps ensured the illuminated slits were not visible to the camera, so the only light to reach the sensor would be that scattered by the sedimenting grains.

\begin{figure}[ht]
\includegraphics[width=\columnwidth]{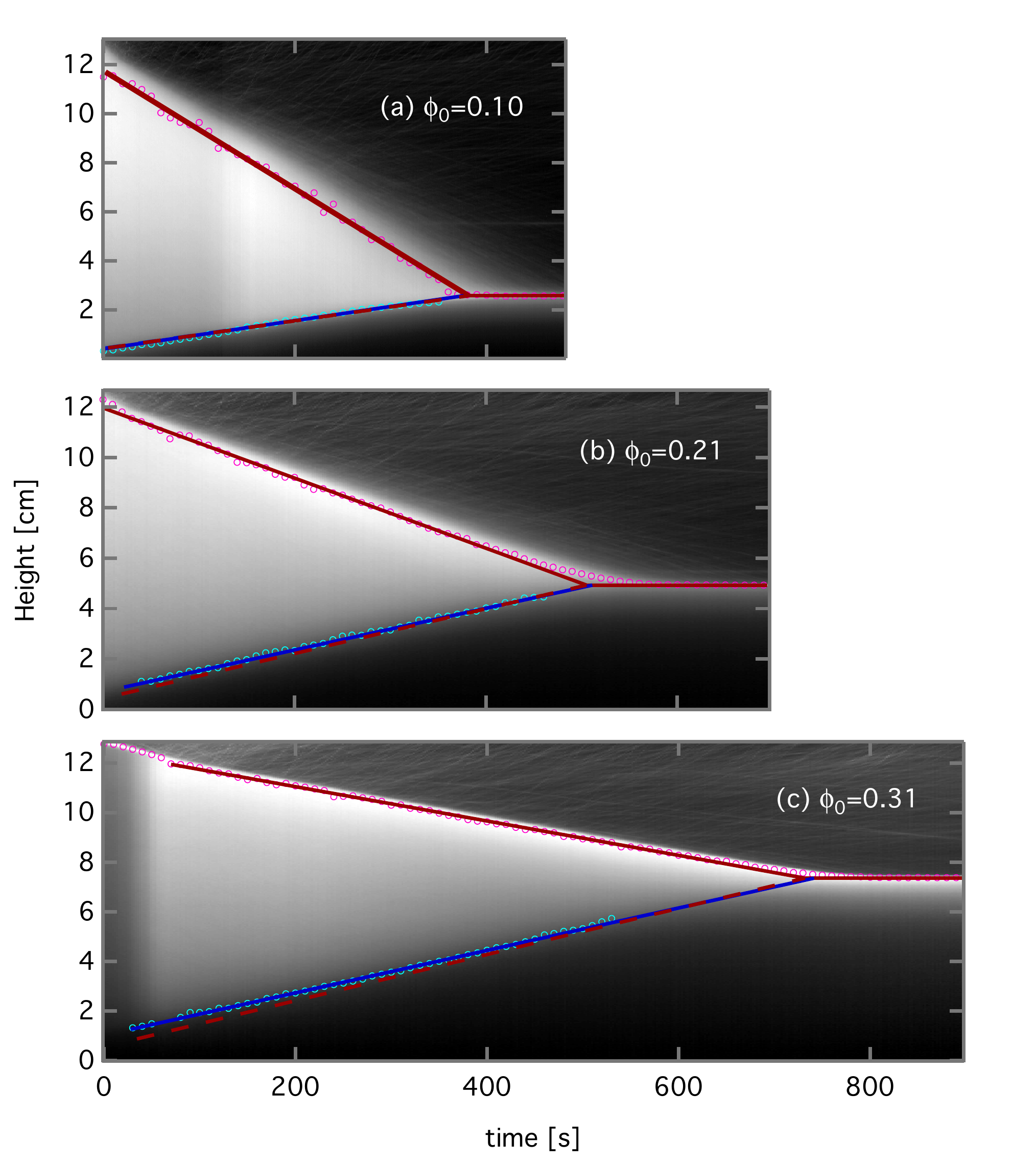}
\caption{Kymographs for sedimentation of $d=365~\mu m$ grains at different initial volume fractions, as labeled, constructed from the central 1~cm wide portion in a time series of images as shown in Fig.~\ref{Montage}.
Height=0 is the bottom of the container.
In all cases, the supernatant appears as a dark region which grows down from the top, the packing as a dark region that grows up from the bottom, and the dispersion as a narrowing bright region in between.
Interface positions (small points) are determined by peaks in vertical intensity gradients.
The uncertainty in location is comparable to symbol size.
The solid lines are linear fits to interface position versus time, giving $v$ and $v_c$ for upper and lower interfaces, respectively.
The expectation for the position of the lower interface based on the $v$ and Eq.~(\ref{eq:vc}) is plotted as a dashed maroon line.}
\label{Kymograph}
\end{figure}

A characteristic time series of images taken in this manner is shown in Fig.~\ref{Montage} for the $d=365~\mu$m grains.
Corresponding kymographs for three different initial volume fractions are shown in Fig.~\ref{Kymograph}.
Several distinct features emerge: The initial dispersion strongly scatters light, and so appears bright white in the image.
The dispersed grains begin to settle under gravity, and a depleted supernatant appears at the top.
Because the supernatant contains no scatterers, it appears dark.
Finally a dense packing accumulates at the container floor.
It is much denser than the initial dispersion, so much more of the light is back-scattered or absorbed, resulting in another dark region.
The volume fraction of the sediment is estimated as $\phi_c=0.54\pm0.01$ based on the height of the sample and the height of the final packing.

Though all interfaces are blurred by multiple light scattering, they are nonetheless sharp enough to be located as the peak in the vertical gradient of the image intensity.
The interface speeds are then found by linear fits of peak location versus time.
Results are given in Table~\ref{Results}.
If the suspension remains at constant volume fraction during sedimentation, then the supernatant-dispersion speed $v$ is constant and equal to the sedimentation speed.
The dispersion-sediment interface speed $v_c$ is also constant and is related to $v$ by volume conservation as
\begin{equation}
	v_c (\phi_c-\phi) = v \phi,
\label{eq:vc}
\end{equation}
As a first check for consistency, dashed lines of slope $v_c$, calculated from Eq.~(\ref{eq:vc}) and the measured values of $v$, are also plotted on Fig.~\ref{Kymograph}.
Both the upper and lower interface lines are in good visual agreement with these expectations from Eq.~(\ref{eq:vc}).
As a second check, we compute the packing fraction of the sediment two ways: based on Eq.~(\ref{eq:vc}) and the measured interface speeds, and from direct measurement of packing height.
The results for both methods, and the three samples, are all consistent with  $\phi_c=0.54\pm0.01$. 
This number agrees with the random-loose packing expectation for slow non-turbulent deposition of particles with static friction coefficient near one~\cite{Menon10}.
Finally the corresponding values of $H$, and the standard deviation $\Delta H$ based on $v$ results for the four diameters, are calculated and presented in Table~\ref{Results} for the three initial volume fractions.
For each $\phi$, the range of $v$ is a factor of roughly 30 for the four particle sizes.
And all four give a consistent value for $H$, to within about 10\%.
Results for average $H$ and $\Delta H$ are shown in plots and are included in the Supplemental data file~\cite{supp}.

\begin{table*}[ht]
\caption{\label{Results} Mean settling speeds $v$, and jamming front speeds $v_c$, as determined from fits to interface positions versus time, for the four grain sizes ($d=180$, 365, 515, and 1000~$\mu$m, ordered left to right).
Also presented are the corresponding values for the average and standard deviation of the hindered settling function $H=v/v_s$.}
\begin{ruledtabular}
\begin{tabular}{cccc}
$\phi$ & $v$ ($\mu$m/s) & $v_c$ ($\mu$m/s)  & $H$  \\ \hline
0.10 & $73.23\pm0.05$, $240.7\pm0.1$, $572.0\pm0.3$, $2044\pm3$ & $18.31\pm0.06$, $57.6\pm0.1$, $137.7\pm0.3$, $396\pm8$ & $0.648\pm0.061$ \\
0.21 & $40.26\pm0.03$, $139.2\pm0.1$, $313.1\pm0.3$, $1211\pm2$ & $23.1\pm0.2$, $82.7\pm0.5$, $173.8\pm0.5$, $687\pm5$ & $0.367\pm0.027$ \\
0.31 & $21.76\pm0.03$, $69.18\pm0.07$, $169.6\pm0.2$, $590\pm1$ & $25.1\pm0.4$, $85.5\pm0.4$, $184.2\pm0.5$, $687\pm5$ & $0.189\pm0.022$
\end{tabular}
\end{ruledtabular}
\end{table*}

\section{Meta-analysis}

\begin{figure}[ht]
\includegraphics[width=\columnwidth]{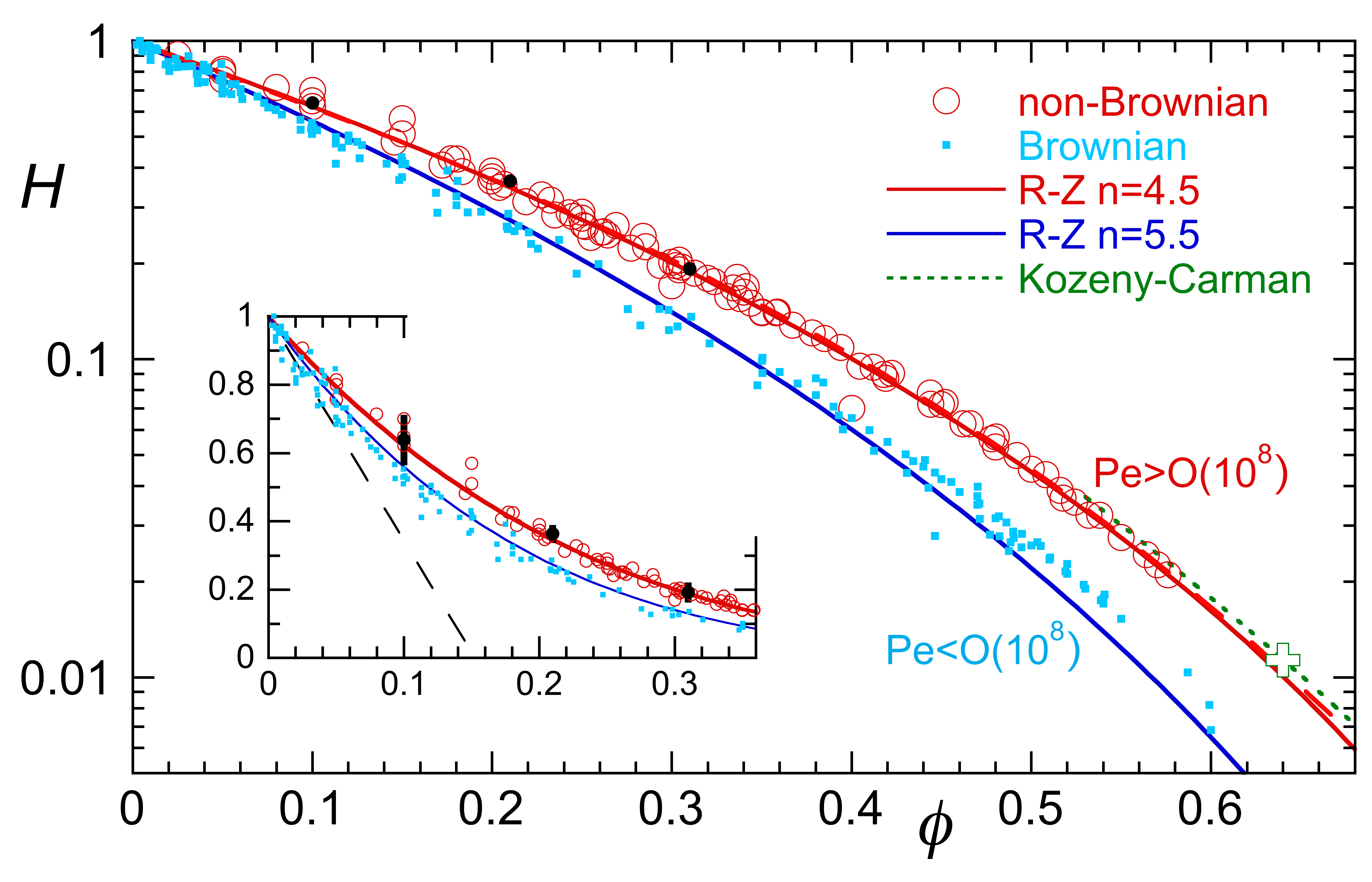}
\caption{Hindered settling function $H=v/v_s$ for suspensions with uniform volume fraction $\phi$, where $v$ is the average settling speed and $v_s$ is the single-grain Stokes settling speed.
The small solid black circles are our data, based on four different particle sizes (Table~\ref{Results}).
All other data are taken from the literature (Table~I and Supplemental Material \cite{supp}):
Open red circles are for larger non-Brownian particles with ${\rm Pe}>\mathcal O(10^8)$ \cite{Richardson54, Oliver61, Ham88, Ham90, Nicolai95}.
Small solid blue squares are for smaller Brownian particles with ${\rm Pe}<\mathcal O(10^8)$ \cite{Buscall82, Kops82, Bacri86, Davis88, Ackerson90, Chaikin92, Salin95, Ackerson07, Piazza08}.
The solid curves are the Richardson-Zaki form $H(\phi)=(1-\phi)^n$ with exponents as labeled.
The dashed curve, nearly indistinguishable from $n=4.5$, is $H=\exp[-4.76\phi-5.75\phi^3]$.
The dotted curve is the Kozeny-Carman form $H=18K=(1-\phi)^3/(10\phi^2)$ for sintered spheres; the value for random close packing at $\phi\rightarrow\phi_c=0.64$ is particularly well-established \cite{Beavers73, Verneuil11} and is shown as an open cross.
The dashed line in the inset is $(1-6.55\phi)$ \cite{Batchelor72}.}
\label{HvsPhi}
\end{figure}

The compiled hindered settling data are all plotted in Fig.~\ref{HvsPhi} versus particle volume fraction.
To our surprise, individual data sets sort cleanly onto two distinct branches: an upper one for the larger non-Brownian particles and a lower one for the smaller Brownian particles.
The non-Brownian branch has less hindering, i.e.~larger $H(\phi)$, and nearly merges smoothly onto the Kozeny-Carman expectation from the permeability of sintered spheres.
This merger, combined with the good data collapse, provides vastly improved confidence in the empirical behavior.
At all $\phi$, including up to close packing, both branches of data are well-described by the Richardson-Zaki form $H(\phi)=(1-\phi)^n$, as shown.
For the larger non-Brownian particles, the fits are excellent and tightly constrain the exponent to $n=4.48\pm0.04$.
For the smaller Brownian particles, the fits are good and constrain the exponent to $n=5.6\pm0.1$.
Fitting procedures are fully discussed in the Appendices.
The Richardson-Zaki forms are plotted with exponents rounded to 5.5 and 4.5, respectively; these bracket the Ref.~\cite{GuazzelliMorris} recommendation of $n\approx 5$ for $\phi<0.4$.

We believe that prior equivocations and contradictions for $H(\phi)$ are resolved in light of there being two branches, previously unrecognized, combined with large uncertainty in fits to $n$ for data sets with limited $\phi$ ranges.
So our compilation is an important advance.
It is further important in establishing the existence of a crossover from Brownian to non-Brownian behavior for increasing P\'eclet number, at ${\rm Pe}_c =\mathcal O(10^8)$.
This rough value is deduced from inspection of Table~\ref{DataComp}, where sets are sorted by Pe and labeled according to the branch on which the data lie.
The Davis88 \cite{Davis88} data set is slightly out of order in this regard: it shows Brownian behavior even though its Pe is a bit larger than two non-Brownian data sets.

The very large value of ${\rm Pe}_c$ is surprising \cite{JohnBrady}, and means that very little Brownian motion is sufficient to affect the hydrodynamic interactions and particle fluctuations/configurations that otherwise occur in non-Brownian sedimentation.
The extreme sensitivity of sedimentation to thermal noise is even greater than that for the reversibility of shear-induced rearrangements \cite{Arratia17}.
In this regard, some authors \cite{Buscall82, Davis88, Chaikin92} have convincingly but incorrectly pronounced suspensions with seemingly large Pe, e.g.~$10^5$, to be non-Brownian.
Likewise, Snabre-Mills \cite{Snabre00} incorrectly state that the Buscall82, Paulin90, and Xue92 data sets are for non-Brownian sedimentation as per their theory.
The extreme sensitivity to thermal fluctuations suggests that other sources of noise could also affect sedimentation.
Ambient vibrations are unlikely to play role, since they must be different for the different experiments yet the data collapse.
Lastly, since existing data are insufficient to capture the crossover between branches and to precisely locate ${\rm Pe}_c$, we hope our work motivates a new generation of sedimentation experiments.

As to {\it why} the Brownian branch should be lower and slower than the non-Brownian branch, we can only speculate.
On the grossest level one may say that Brownian motion helps keep the particles suspended.
Microscopically, perhaps Brownian motion makes the permeation flow less smooth/steady and hence more dissipative/slow.
It is also likely that thermal energy helps separate contacting particle pairs that would otherwise sediment faster than single grains.
This latter possibility is consistent with simulation results with~\cite{LaddJCP90} and without~\cite{LaddPF93} Brownian motion.
It remains a challenge to firm up such intuition in terms of a theoretical prediction for the value of ${\rm Pe}_c$, which could conceivably depend on $\phi$.


\section{Conclusion}

By comparing new measurements plus extensive prior measurements from the literature \cite{supp}, we have shown that hindered settling function data for monodisperse spheres at low Re are consistent with one of two behaviors.
Smaller Brownian particles experience stronger hindering, and thus settle at a smaller fraction of their Stokes velocity than larger non-Brownian particles at the same volume fraction.
The two branches are described well by hindered settling functions of the Richardson-Zaki form~\cite{Richardson54}, $H(\phi)=(1-\phi)^{n}$, with exponents of $n=5.6\pm0.1$ for Brownian particles and $n=4.48\pm0.04$ for non-Brownian particles.
This holds from the dilute limit all the way up to close packing.
The crossover between the two branches happens at a surprisingly large P\'eclet number, $\mathcal O(10^8)$ meaning that the sedimentation speed is slowed by surprisingly little Brownian motion.
These findings are important for accurately establishing the empirical behavior over the whole range of volume fractions, for eliminating equivocations and contradictions in previous publications, and for warning that thermal effects are more important than usually thought.
This sharpens the challenge to understand the origin of the Richardson-Zacki form, and raises new challenges to understand the values of the two exponents as well as the surprising location and quickness of the crossover between the two branches.

\begin{acknowledgments}
We thank P. Arratia, J. Brady, M. Brenner, E. Guazzelli, G. Homsy, A. Ladd, T. Lubensky, J. Martin, J. Morris, D. Pine, D. Salin, P. Tong, S. Torquato, D. Weitz, and P.-Z. Wong for helpful conversations.  This work was supported by the NSF through grant numbers DMR-1305199, DMR-1619625, MRSEC/DMR-112090, and MRSEC/DMR-1720530.
\end{acknowledgments}


\appendix

\section{The non-Brownian Branch}

\begin{figure}[ht]
\includegraphics[width=\columnwidth]{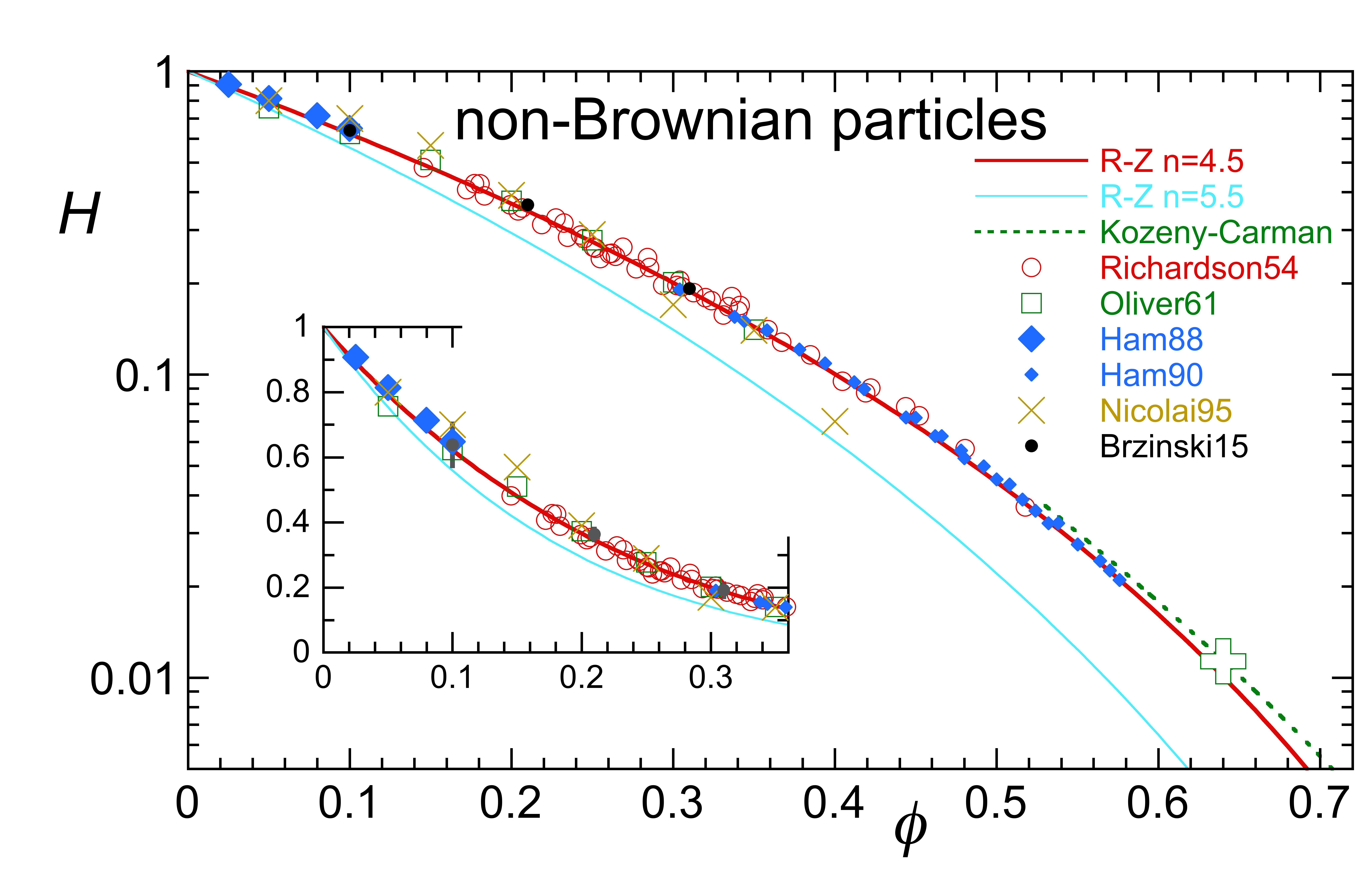}
\caption{(color online) Hindered settling function versus volume fraction for non-Brownian particles. Individual data sets are specified per Table~\ref{DataComp}.}
\label{Hnonbrownian}
\end{figure}

In this appendix and the next, we plot the individual data sets within each branch and fit them carefully to different forms.
We begin with the hindered settling function for the all the non-Brownian data, shown in Fig.~\ref{Hnonbrownian}.
Close inspection reveals how the individual data sets appear mutually consistent and randomly scattered around the plotted $n=4.5$ Richardson-Zaki form.
However there is one exception: the data labeled Nicolai15 from Table~I of Ref.~\cite{Nicolai95} smoothly rise above and then dip below the other non-Brownian data with increasing $\phi$.
Ironically this is the major data set highlighted in Ref.~\cite{GuazzelliMorris} that led to the equivocal recommendation of $n\approx 5$.
We also point out that every data point in Fig.~\ref{Hnonbrownian} represents one measurement, with one exception: each of our own data points represents measurements from four different particle sizes.

To analyze the data we fit to three hindered settling functions:  Richardson-Zaki and Snabre-Mills, in Eq.~(\ref{Hexpect}), plus a cumulant expansion form $H(\phi)=\exp[-n\phi-n_2\phi^2-n_3\phi^3+\mathcal O(\phi^4)]$.
The rationale for the latter is that the data bend downward gradually (not dramatically) on a semi-logarithmic plot; this function expands as $H(\phi)=1-n\phi+\mathcal O(\phi^2)$ like Richardson-Zaki, while Snabre-Mills expands as $1-b\phi+\mathcal O(\phi^2)$.
The fits are performed using the tabulated uncertainties $\Delta H$ as weighting.
The Nicolai95 \cite{Nicolai95} data set is excluded, since it deviates systematically from the other non-Brownian data sets.
The fitting results are as follows:
\begin{widetext}
\begin{equation}
	h(\phi) =
	\begin{cases}
		(1-\phi)^{4.46\pm0.01}  & \text{$\chi^2=1.016,~R=0.99874$} \cr
		\exp[-(4.80\pm0.13)\phi+(0.21\pm0.63)\phi^2-(6.00\pm0.76)\phi^3] & \text{$\chi^2=1.027,~R=0.99873$} \cr
		\exp[-(3.86\pm0.04)\phi-(4.71\pm0.10)\phi^2] & \text{$\chi^2=1.304,~R=0.99788$} \cr
		\exp[-(4.76\pm0.02)\phi-(5.75\pm0.11)\phi^3] & \text{$\chi^2=1.028,~R=0.99872$} \cr
		b=3.61\pm0.01 & \text{$\chi^2=2.508,~R=0.99688$} \cr
	\end{cases}
\label{NBfits}
\end{equation}
\end{widetext}
where $\chi^2 = \langle \{[H_{data}(\phi)-H_{fit}(\phi)]/\Delta H\}^2\rangle$.
Note that the cumulant expansion is evidently better with the $\phi^2$ term set to zero.
To check the quality of these forms, we also fit over only the first half of the compilation ($\phi\le0.305$), giving:
\begin{widetext}
\begin{equation}
	h(\phi) =
	\begin{cases}
		(1-\phi)^{4.49\pm0.03}  & \text{$\chi^2=0.998,~R=0.99589$} \cr
		\exp[-(4.22\pm0.31)\phi-(3.62\pm2.97)\phi^2-(0.78\pm6.85)\phi^3] & \text{$\chi^2=0.945,~R=0.99611$} \cr
		\exp[-(4.19\pm0.14)\phi-(3.96\pm0.55)\phi^2] & \text{$\chi^2=0.946,~R=0.99611$} \cr
		\exp[-(4.59\pm0.09)\phi-(8.98\pm1.27)\phi^3] & \text{$\chi^2=0.977,~R=0.99598$} \cr
		b=3.92\pm0.03 & \text{$\chi^2=1.002,~R=0.99564$} \cr
	\end{cases}
\label{NBfits2}
\end{equation}
\end{widetext}
Over both ranges, the Richardson-Zaki form produces the best fit with a combined estimate of $n=4.48\pm0.04$ for the exponent.
While this works well, the cumulant form with the $\phi^2$ term set to zero is nearly indistinguishable, and more likely to be explained by theory.
These two forms give slightly different leading behavior, which we estimate more conservatively as $H(\phi)=1-(4.5\pm0.2)\phi$.
This is the exponent used for the Richardson-Zaki form shown in all plots.
Forcing the Richardson-Zaki exponent to $n=4.5$ gives a ``fit'' with goodness $\chi^2=0.993$ and ${\rm R}=0.99838$, and describes the data well from $\phi=0$ all the way up to essentially $\phi_c$.
The form $H(\phi)=\exp[-4.76\phi-5.75\phi^3]$ is nearly identical to within the level of scatter in the data.

\section{The Brownian Branch}

\begin{figure}[ht]
\includegraphics[width=\columnwidth]{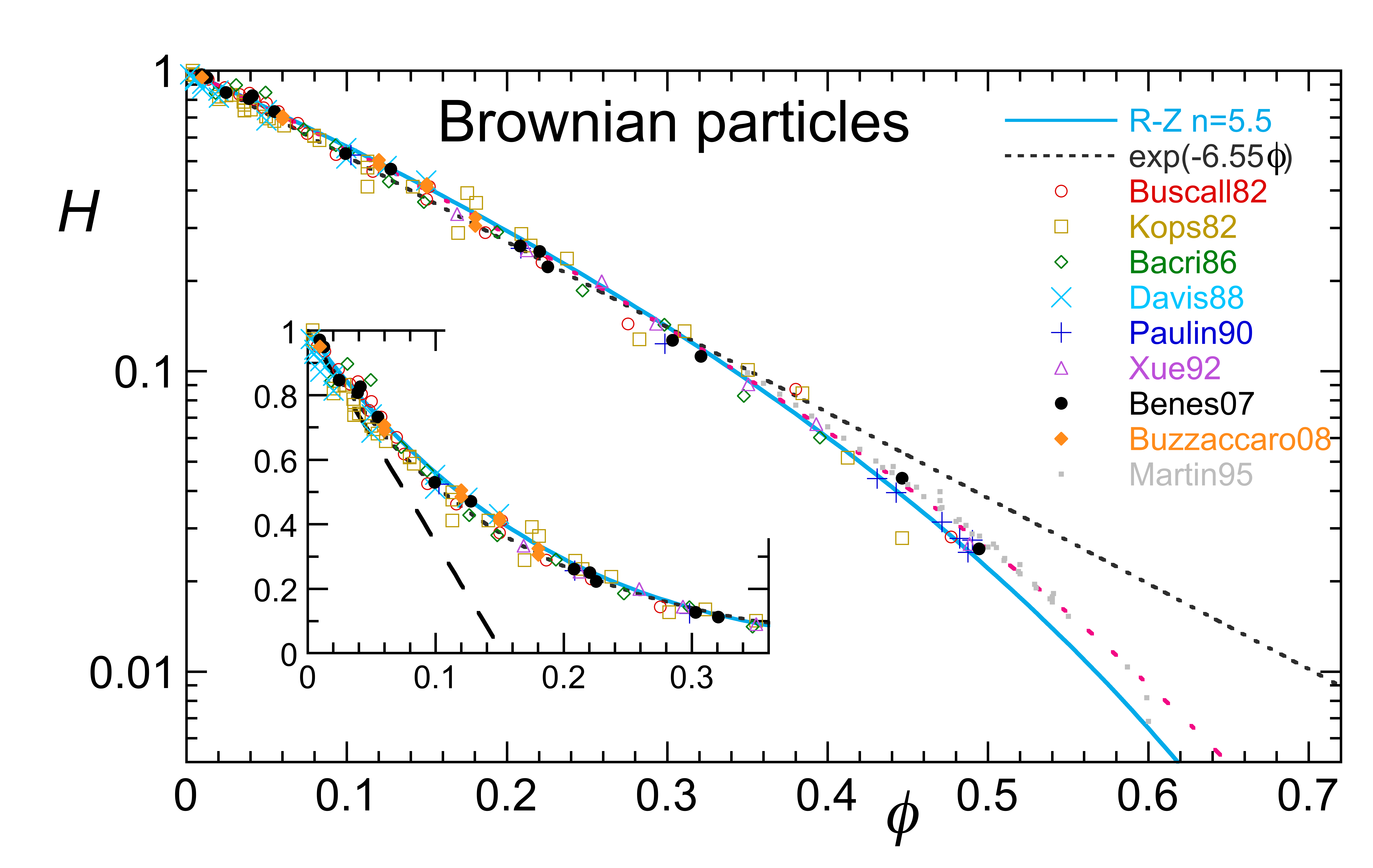}
\caption{(color online) Hindered settling function versus volume fraction for Brownian particles. Individual data sets are specified per Table~\ref{DataComp}.
The dashed line in the inset represents Batchelor's $H=1-6.55\phi$ prediction for the leading behavior \cite{Batchelor72}.
The sparsely dotted curve is for the Snabre-Mills form with the authors' recommended value of $b=5.6$ \cite{Snabre00}.}
\label{Hbrownian}
\end{figure}

Hindered settling data for the Brownian case is shown in Fig.~\ref{Hbrownian}.
There are more data sets than for the non-Brownian case, and none appear by eye to deviate from the average trend.
As above, good fits can be also obtained to the various forms.
However, analysis is more difficult in that different functions and fitting ranges give significantly different leading behavior:
\begin{widetext}
\begin{equation}
	h(\phi) =
	\begin{cases}
		(1-\phi)^{5.53\pm0.02}  & \text{$\chi^2=1.054,~R=0.99699$} \cr
		\exp[-(5.92\pm0.19)\phi-(1.50\pm1.33)\phi^2-(3.27\pm2.07)\phi^3] & \text{$\chi^2=0.843,~R=0.99759$} \cr
		\exp[-(5.65\pm0.08)\phi-(3.58\pm0.22)\phi^2] & \text{$\chi^2=0.864,~R=0.99753$} \cr
		\exp[-(6.13\pm0.06)\phi-(5.55\pm0.34)\phi^3] & \text{$\chi^2=0.854,~R=0.99756$} \cr
		\exp[-6.55\phi+(2.64\pm0.40)\phi^2-(9.28\pm0.92)\phi^3] & \text{$\chi^2=0.929,~R=0.99735$} \cr
		b=5.84\pm0.05 & \text{$\chi^2=0.874,~R=0.99749$} \cr
	\end{cases}
\label{Bfits}
\end{equation}
\end{widetext}
The first of these fits is close to $(1-\phi)^{5.4}$ shown in Fig.~10 of Ref.~\cite{Gilleland11} along with the Buscall82, Paulin90, and Xue92, and data sets.
The penultimate of these fits has leading behavior set to Batchelor's prediction $H(\phi)=1-6.55\phi$; this gives a good fit too.
Snabre-Mills works well for the Brownian branch, but not the non-Brownian.
All these fits, above and below, were done before we knew of the Martin95 data set.
Performing the fits over a restricted range ($\phi\le0.305$) gives
\begin{widetext}
\begin{equation}
	h(\phi) =
	\begin{cases}
		(1-\phi)^{5.71\pm0.04}  & \text{$\chi^2=0.658,~R=0.99544$} \cr
		\exp[-(6.04\pm0.36)\phi+(0.46\pm3.90)\phi^2-(9.68\pm9.91)\phi^3] & \text{$\chi^2=0.648,~R=0.99551$} \cr
		\exp[-(5.73\pm0.16)\phi-(3.28\pm0.75)\phi^2] & \text{$\chi^2=0.657,~R=0.99544$} \cr
		\exp[-(6.00\pm0.10)\phi-(8.53\pm1.92)\phi^3] & \text{$\chi^2=0.648,~R=0.99551$} \cr
		\exp[-6.55\phi+(5.71\pm1.05)\phi^2-(22.2\pm4.3)\phi^3] & \text{$\chi^2=0.547,~R=0.99537$} \cr
		\exp[-(6.41\pm0.05)\phi] & \text{$\chi^2=0.893,~R=0.99420$} \cr
		b=5.78\pm0.07 & \text{$\chi^2=0.665,~R=0.99565$} \cr
	\end{cases}
\label{Bfits3}
\end{equation}
\end{widetext}
Overall the Richardson-Zaki form with $n\approx5.5$ can thus be taken as a good empirical description of behavior for the entire Brownian hard sphere compilation.
This is what is shown in all plots.
Snabre-Mills is just as good if not better, which is somewhat ironic since it was intended for {\it non}-Brownian samples.
However, neither form can be said to capture the leading behavior.
Rather, from the increase of $n$ with restricted fitting range, and the results of the last two fits, it appears that the leading behavior is consistent with Batchelor's prediction.
Furthermore, as seen in Fig.~\ref{Hbrownian}, the simple form $H(\phi)=\exp(-6.55\phi)$ gives a fine description for $\phi<0.4$.
But a different picture emerges from fitting over all $\phi$ for only the four most recent data sets, which have the least scatter:
\begin{widetext}
\begin{equation}
	h(\phi) =
	\begin{cases}
		(1-\phi)^{5.50\pm0.03}  & \text{$\chi^2=0.985,~R=0.99818$} \cr
		\exp[-(5.58\pm0.26)\phi-(3.60\pm1.18)\phi^2-(0.26\pm2.70)\phi^3] & \text{$\chi^2=0.584,~R=0.99892$} \cr
		\exp[-(5.56\pm0.10)\phi-(3.77\pm0.25)\phi^2] & \text{$\chi^2=0.585,~R=0.99892$} \cr
		\exp[-(6.08\pm0.07)\phi-(5.66\pm0.39)\phi^3] & \text{$\chi^2=0.687,~R=0.99873$} \cr
		\exp[-6.55\phi+(2.82\pm0.48)\phi^2-(9.54\pm1.10)\phi^3] & \text{$\chi^2=0.937,~R=0.99827$} \cr
		b=5.79\pm0.06 & \text{$\chi^2=0.684,~R=0.99874$} \cr
	\end{cases}
\label{Bfits2}
\end{equation}\end{widetext}
Here, the leading behavior in the first three fits is actually consistent with $n=5.6\pm0.1$ and not Batchelor.  Further hindered settling data for colloidal Brownian spheres would be helpful.

\bibliography{SedRefs}

\begin{thebibliography}{41}
\bibitem [{Joh()}]{JohnBrady}%
  \BibitemOpen
  \href@noop {} {}\bibinfo {note} {J. F. Brady suggests that the
  extraordinarily large P\'eclet number at the crossover between Brownian and
  non-Brownian behavior is not so alarming if considered in terms of particle
  radius: ${\rm Pe}_c = 4\pi\Delta\rho g a_c^4/3k_bT \approx 10^8$ gives $a_c
  \approx 70(k_BT/\Delta\rho g)^{1/4}$.}\BibitemShut {Stop}%
\end{thebibliography}

\end{document}